\documentclass[12pt,preprint]{aastex}
\usepackage{emulateapj5}

\newcommand{\hi}	{\ion{H}{1}}
\newcommand{\kms}	{km~s$^{-1}$}
\newcommand{\surfb}	{mag~arcsecond$^{-2}$}
\newcommand{\rs}	{R$_{\mbox{s}}$}
\newcommand{\rsm}	{R_{S}}
\newcommand{\sbr}	{\mu_{\mbox{{\tiny R}}}}
\newcommand{\sbv}       {\mu_{\mbox{{\tiny V}}}}
\newcommand{\simgtr}    {\; \raisebox{-.2ex}{$\stackrel{>}{\mbox{\tiny
$\sim$}}$} \;}
\newcommand{\simless}   {\; \raisebox{-.2ex}{$\stackrel{<}{\mbox{\tiny
$\sim$}}$} \;}

\slugcomment{Submitted for publication to \emph{The Astrophysical
Journal}}

\shorttitle{Absence of Stars in Compact HVCs}
\shortauthors{Simon \& Blitz}

\begin{document}

\title{The Absence of Stars in Compact High-Velocity Clouds}
\author{Joshua D. Simon and Leo Blitz}
\affil{Department of Astronomy, University of California at Berkeley}
\affil{601 Campbell Hall, Berkeley, CA  94720}
\email{jsimon,blitz@astro.berkeley.edu}

\begin{abstract}

We present the results of our search for faint Local Group dwarf
galaxies in compact high-velocity clouds (HVCs).  We used digitized
Palomar Observatory Sky Survey (POSS) data to examine 1 square degree
of sky around each of $\sim250$ northern hemisphere HVCs.  The POSS
images were processed to remove foreground stars and large-scale
backgrounds, smoothed to enhance arcminute-sized low surface
brightness features, and then compared to the original plates.  Using
this technique, we located 60 candidate dwarf galaxies in the
$\sim250$ degrees$^{2}$ that we surveyed.  Followup observations of
these candidates have revealed several distant clusters of galaxies
and a number of possible Galactic cirrus clouds, but no Local Group
dwarfs.  It appears that many of the low surface brightness features
in the sky survey data are plate flaws.  The second-generation red
POSS plates are sensitive down to surface brightness levels of $25-26$
magnitudes arcsecond$^{-2}$, and our followup images reached
$10\sigma$ limiting magnitudes of $R=21-23$ for point sources.  Given
these limits, all known Local Group galaxies except four of the very
diffuse, extended dwarf spheroidals located within 100 kpc of the
Milky Way would have been detected had they been in our survey.
Therefore, we can rule out the possibility that these HVCs are
associated with normal but faint dwarf galaxies.  If compact HVCs do
contain stars, they must have surface brightnesses $\simgtr 1$ \surfb\
fainter than most known Local Group galaxies.

\end{abstract}

\keywords{galaxies: dwarf --- intergalactic medium --- Local Group}

\section{INTRODUCTION}

The nature of the high-velocity clouds (HVCs) of neutral hydrogen has
been a source of controversy since their discovery almost 40 years
ago and remains so today.  It now appears that HVCs do not represent
a single phenomenon.  Rather, they are an amalgam of several types of
objects that were grouped together by the extremely broad
observational definition of ``HVC'': neutral hydrogen that is moving
at velocities inconsistent with simple models of Galactic rotation.
Some HVCs, notably the Magellanic Stream, are composed of gas that has
been either tidally or ram-pressure stripped out of the Magellanic
Clouds.  Others are very extended clouds close to the Milky Way
(distances of a few to a few tens of kpc), whose origin is not clear,
although they appear to be extragalactic.  Finally, there is a large
population of small clouds (in angular size) whose properties have
proven very difficult to determine.  These were labeled compact HVCs
by \citet[hereafter BB99]{bb99} and may represent a relatively
homogeneous class of objects.  We will concentrate on these HVCs for
the remainder of this paper.

One hypothesis that has received attention is that HVCs are clouds of
primordial gas in the Local Group (LG), and thus are located at
typical distances from the Milky Way (MW) of up to 1 Mpc
\citep{oort66,oort70,verschuur69,ks69}.  \citet[hereafter B99]{b99}
have recently proposed a dynamical model based on this idea that
simulates the evolution of the Local Group. The model is able to
explain the properties of HVCs as the postulated primordial clouds,
left over from the formation of the Local Group and currently falling
into the LG barycenter.  In addition to reproducing observed results
of the spatial distribution and kinematics of HVCs, B99 make
observational predictions about HVC internal pressures, metallicities,
and H$\alpha$ emission, all of which are being actively investigated.

Other potentially viable theories to explain the HVC puzzle include
the Galactic Fountain model \citep{sf76,bregman80} and the Tidal
Debris model.  According to the Galactic Fountain model, large numbers
of supernova explosions in the inner Galaxy expel hot, metal-rich gas
from the Milky Way's disk into the halo.  These clouds then cool and
fall back towards the Galaxy and are seen as HVCs.  In contrast, the
Tidal Debris model proposes that all HVCs are remnants of the tidal
interaction between the Milky Way and the Magellanic Clouds.

The best way to settle the question of the origin of HVCs is to
measure their distances, since the three models prefer distances that
differ by two orders of magnitude.  In this paper, we describe our
search for stars in HVCs, which if present could be used for
photometric distance estimates.  Despite the large amount of
observational effort at 21-cm that has been devoted to HVCs, there
have not yet been any systematic, large-scale searches for stars.

In Section 2, we will discuss the motivations for this work in more
detail.  Readers who are already familiar with the recent HVC
literature may wish to skip ahead.  We will describe our search
technique and discuss the datasets used in our analysis in Section 3.
Section 4 will briefly present the results of our search of the
Palomar Observatory Sky Survey (POSS) data.  In Section 5, we will
describe our followup observations.  The limits we have placed on the
stellar content of HVCs and the implications of our results will be
discussed in Section 6, and our conclusions will be presented in
Section 7.

\section{MOTIVATIONS}
\subsection{Recent HVC Results}

Recent observations of HVCs have tended to at least roughly agree with
the predictions made by B99.  The internal pressures of high-velocity
clouds, as determined by ultraviolet absorption line studies, are low
in the few clouds that have been observed.  This finding is consistent
with the HVCs being LG objects \citep{sembach99}.  The metallicities
of HVCs (other than the Magellanic Stream material) have been found to
be 0.1 -- 0.3 times solar, also within the range predicted by B99
\citep{wakker99,murphy00,richter01,gibson01}.  Some of these
measurements are toward the high end of what can be reasonably
accommodated by the Blitz et al. model, and if lower metallicity HVCs
are not found, this may eventually present a problem for the
extragalactic hypothesis.  The Galactic Fountain model, though, faces
an even bigger challenge, since it predicts solar or higher
metallicities for high-velocity gas.  It should be noted that all of
these metallicity determinations are for nearby HVCs that are part of
the large complexes.  These complexes are believed to be interacting
with the Milky Way, and might have been partially enriched as a
result.  Therefore, these metallicity measurements do not necessarily
apply to the presumably more distant and isolated compact clouds, and
in fact, may be regarded as upper limits to the metallicities of
compact HVCs.  A further caveat to the metallicity determinations is
that they rely on comparing optical and UV absorption column densities,
which probe gas on scales of $\sim 10^{-4}$ arcseconds, with \hi\
observations on an angular scale at least four orders of magnitude
larger.  The effects of this mismatch are not known.

H$\alpha$ emission from HVCs has also been observed, but the results
of this work are difficult to interpret.  A number of researchers have
used Fabry-Perot instruments to detect $\sim20$ HVCs in H$\alpha$, and
sometimes also in the nearby \ion{N}{2} and \ion{S}{2} emission lines
\citep{kutyrev89,tufte98,bh01,weiner01}.  By comparing the strength of
the emission from these clouds with that from gas at known distances
(e.g., nearby complexes or the Magellanic Stream), one can estimate
the distances to the HVCs.  \citet{bh01} and \citet{weiner01} conclude
based on their observations that the HVCs they detect are probably
nearby, at distances of tens of kpc.  However, their method depends
critically on the assumption that HVC ionization is caused primarily
by photoionization from the Milky Way's UV field.  Although this
assumption seems reasonable, it appears to be false for the Magellanic
Stream gas \citep{ww96}.  Furthermore, these calculations do not work
well for one object whose distance is known: the Sculptor dwarf
spheroidal galaxy.  Sculptor is located 80 kpc from the Milky Way, yet
it is brighter in H$\alpha$ than Complexes A, C, and M, which are all
an order of magnitude closer.  If the HVC distance estimates are
normalized to Sculptor instead of Complexes A and M, significantly
larger distances (up to $\sim200$ kpc) will be derived.  In support of
the shorter distance scale, \citet{weiner01} note that none of the
HVCs they observe have fluxes close to their detection limit; every
object is at least an $8\sigma$ detection.  They argue that the lack
of H$\alpha$-faint HVCs implies that HVCs are not very far away.  As
with the metallicity studies cited above, though, many of these
observations have been of the HVC complexes that were already known
or suspected to be nearby.  Until there has been a comprehensive
H$\alpha$ survey of the compact HVCs {\it and} there is a reliable way
to relate HVC H$\alpha$ fluxes to distances, these observations will
not offer strong confirmation or refutation of any HVC models.

\subsection{A New Technique}

Faced with this confused state of affairs, we decided to take a
different approach to the HVC problem.  The key objects are the ones
whose nature is most uncertain: compact HVCs.  Since there does not
appear to be a simple way to determine their distances directly with
current techniques, we have attempted an indirect method --- searching
for stellar counterparts to the HVCs.  If we could detect such stars,
it would be straightforward to obtain accurate photometric distances
to them, and thereby conclusively establish the nature of the
high-velocity clouds.

Several lines of evidence have recently led to the hypothesis that
HVCs are likely to contain stars.  First, blind \hi\ surveys of nearby
galaxies and groups have found that the low-mass \hi\ clouds they
detect are often associated with low-surface brightness (LSB) dwarf
galaxies \citep{banks99,k-k99,pw99,rs00,boyce01}.  This suggests that
the LG \hi\ clouds might be similarly associated with currently
undiscovered dwarf galaxies.  Second, \hi\ has now been detected
towards half of the LG dwarf spheroidal (dSph) galaxies and all of the
dwarf irregular (dIrr) galaxies \citep{br00}.  In one case, this \hi\
may have been previously classified as an HVC (HVC 561 = Sculptor) by
\citet{wvw}.  The \hi\ around the dSphs appears to be very extended
compared to the optical galaxies and is of similar physical size and
mass to the HVCs as described by B99.  These similarities support the
idea that HVCs and \hi\ in dSphs could be related objects.

A further reason to suspect that HVCs might be associated with dwarf
galaxies comes from numerical simulations of Cold Dark Matter (CDM)
cosmological models.  It has become well-known that these simulations
produce too much substructure \citep{klypin99,moore99}.  In simulated
Local Groups, the mass concentrations around MW and M31 analogues
contain up to a few hundred small dark matter halos.  The real LG is
comparatively barren, with only $\sim35$ known dwarf galaxies.  If CDM
is correct and these dark matter minihalos exist, they probably would
have accreted some gas, and then could subsequently have formed stars
(although see \citet{bullock00} for one explanation of why star
formation might be suppressed in such objects).  In any case, it has
been suggested by several authors that compact HVCs could be these
``missing'' dark matter halos (e.g., \citet{klypin99,moore99}).

Finally, it is almost certain that the census of Local Group galaxies
is not yet complete.  In the past few years, four new dwarfs have been
discovered: And V by \citet[hereafter A98]{a98}, And VI by
\citet[hereafter KK99]{kk99} and \citet{armandroff99} independently,
And VII by KK99, and Cetus by \citet{whiting99}.  A fifth dSph,
Antlia, was first cataloged by \citet{corwin85} and \citet{fg85}, and
detected in \hi\ by \citet{fouque90}, who suspected it to be a Local
Group member.  Followup photometry confirming this conjecture was not
acquired until the galaxy was rediscovered by \citet{whiting97}
several years later.  In addition to these recent discoveries,
\citet{mateo98} notes that there is an apparent deficit of dwarf
galaxies at low Galactic latitudes (relative to a uniform distribution
on the sky).  On this basis, he postulates up to 15 -- 20 LG dwarfs
remaining to be discovered at $b \leq 30\degr$.

\section{SEARCH METHODOLOGY}

\subsection{POSS Image Processing}

The machinery for our search is based loosely on the work of A98 for
their similar project (a blind survey for dSph companions of M31).
Starting from the outline they provided, we developed an algorithm
that processes digitized POSS images\footnote{The Digitized Sky Survey
was produced at the Space Telescope Science Institute under
U.S. Government grant NAG W-2166. The images of these surveys are
based on photographic data obtained using the Oschin Schmidt Telescope
on Palomar Mountain and the UK Schmidt Telescope. The plates were
processed into the present compressed digital form with the permission
of these institutions.} from the Space Telescope Science Institute
(STScI) to enhance extended LSB objects of the appropriate angular
size to be Local Group dwarf galaxies.  We adopt similar steps, but in
a different order and with significant changes in implementation from
A98.  The algorithm is described in detail in the following
paragraphs.  All of the image processing was done in the IDL
environment.

We first attempt to remove the brightest stars from the image.
Because these stars have such extensive wings, this task will be
incompletely successful, at best.  Nevertheless, we can remove enough
of the flux that these stars will not dominate the smoothed image,
even though they will still be visible.  Bright stars are located with
a simple count threshold --- any pixel with a value of more than 23000
counts was defined to be part of a bright source.  This method can
succeed because our search largely employed the POSS-II red plates
(for $\sim80$\% of our targets), which tend to have similar sky
background levels ($5000 - 6000$ counts on average) and sensitivities.
Having found the brightest stars, we create a circular mask around
each one and replace the pixels in the mask with an average of the
surrounding pixels.

This process is repeated in order to remove the fainter stars from the
image as well.  However, because objects other than faint stars can
have peak brightnesses as high or higher than those of the stars, we
must use a slightly different technique to detect these stars.  We
pass a small ($\approx 10\arcsec$) median filter over the image, and
then subtract the filtered version from the original.  Point sources
are seriously degraded in flux in the filtered image, and so appear
bright in the difference image.  Extended sources are only minimally
affected by the filtering and therefore are largely removed by
subtracting the images.  Objects brighter than a threshold value of
1000 counts per pixel in the difference image are likely to be
foreground stars.  The point sources selected with this criterion are
masked out and replaced, as described above.

Now that the image is relatively free of stellar light, we must deal
with the other major contaminant, the background.  Unfortunately, the
digitized POSS plates are not flat, especially near the edges, and
these background variations significantly hamper a search for LSB
galaxies.  In order to detect the small increases in surface
brightness over a limited area that are associated with dwarf
galaxies, the vignetting of the plates and any other large-scale flaws
in them must be removed as accurately as possible.  We experimented
with two-dimensional polynomial fits to the background, but determined
that the most reliable means of subtracting just the background (and
leaving objects of interest alone) was simply to employ a very wide
median filter.  The filter must be large enough not to pick up
significant signal from dwarf galaxies, or else this procedure could
become counterproductive.  The largest of the distant dSphs are
$\sim5$\arcmin\ across, so a filter more than 10\arcmin\ wide will be
at least four times as large as any dwarf galaxies it encounters.
Empirically, this level of filtering seemed to be safe, so we chose a
filter width of 600 pixels, corresponding to $10.1 \arcmin$ on POSS-II
and Equatorial Red plates, and $17 \arcmin$ on POSS-I and SERC-J
plates.  To deal with the 300 pixel border around the image where the
median filter was too close to the edge to function, we simply
replicate the last line or column that the median filter did produce
outward to the edges of the image.

After removing the background, we treat the bright stars again,
because they still tend to be the brightest sources left in the image
(although not overwhelmingly so).  At every location where a bright
star was removed earlier, we check to see if the average brightness is
still more than half a standard deviation above the median for the
image.  If so, we cut out a larger mask around these remnants, and
again replace it with an average of the surrounding pixels.

Finally, we smooth the image with a dwarf-galaxy-sized filter in order
to further dim the remaining stellar light, and thereby enhance the
relative contrast between high- and low-surface brightness objects.
We use a Gaussian filter here rather than a median filter because the
filter size is such that a median filter runs the risk of smoothing
away the entire image.  According to \citet{davies94}, to optimally
detect LSB objects the filter should be roughly the same size as the
object.  At the distances we expect for the HVCs --- 300 kpc to 1 Mpc
--- dwarf galaxies have angular sizes between 1\arcmin\ and 6\arcmin.
Therefore, we use a 70 pixel ($= 71$\arcsec on POSS-II images and
119\arcsec\ on POSS-I images) filter.  We choose a filter on the small
end of the dwarf galaxy size range because 1) most of the detected
objects will be at large distances (due to the greater survey volume
there), and 2) we do not want to degrade potential dwarf galaxy
signals by smoothing over too large a scale.  Since we are not using
adaptive filtering here, we are not sensitive to LSB sources of all
possible angular sizes.  Objects which are smaller than $\approx
30$\arcsec\ may be missed by our search because of this final filter,
and objects larger than $\approx 15$\arcmin\ could be removed by our
background filter.

The free parameters in this algorithm (order of steps, mask and filter
sizes, etc.) were tuned by extensive testing on known dwarf spheroidal
galaxies, primarily the 6 eponymous companions of Andromeda.  In
Figure 1, we compare processed images of two of the lowest surface
brightness LG dSphs with the original POSS-II plates.  It is clear
that the algorithm works well.  Even though the dwarfs are barely
visible in the unprocessed POSS-II images, they are the brightest
objects in the field of view --- dominating the 10th magnitude
foreground stars --- after smoothing.

\subsection{HVC Catalogs}

We used the catalog of compact high-velocity clouds provided by
\citet{bb99} as the basis for the first part of our search.  They
selected HVCs visually from the Leiden/Dwingeloo Survey of Galactic
Neutral Hydrogen \citep{dap} and from the large HVC catalog compiled
by \citet{wvw}.  BB99 identified a subset of 65 clouds that are both
small (less than $2\degr$ in diameter) and isolated from neighboring
emission.  They claimed that these objects represent a distinct class
of HVCs and are all at Local Group distances.  We included 59 of
these HVCs that were located outside the zone of avoidance ($|b| >
5\degr$) in our sample.  BB99 also presented $\sim 15$ degree$^{2}$
\hi\ maps around each of the compact HVCs (see their Figure 1).  In many
of these moment maps, there are nearby clouds that have comparable
intensities.  Since it is not entirely clear why these clouds were not
included in their catalog, we have chosen 17 of the most prominent of
them to use in our analysis as well, for a total of 76 HVCs from BB99.

In addition to these compact HVCs, we also selected targets from the
new catalog by Robishaw \& Blitz (in preparation, hereafter RB02), a
complete sample derived from an automated search of the
Leiden/Dwingeloo Survey (LDS).  Since this catalog is a work in
progress, we were forced to take a fairly simplistic approach to
defining HVCs.  We first selected all \hi\ detections at velocities
less than $V_{\mbox{\tiny{LSR}}} = -200$ \kms\ or greater than
$V_{\mbox{\tiny{LSR}}} = 200$ \kms.  From this list, we grouped the
\hi\ detections into clouds by defining as part of a single object all
\hi\ within $3\degr$ in both longitude and latitude, and within 10
\kms\ in velocity of the brightest point (the brightest point was also
defined as the central position).  We then culled this list to
eliminate HVCs that could not be observed easily from Lick Observatory
(clouds outside the range $-10\degr \leq \delta \leq 65\degr$).  This
left us with 201 HVCs.  It should be noted that, while the final RB02
catalog will be a flux-limited sample down to the sensitivity limit of
the LDS, this preliminary version is not.  We have certainly missed a
number of HVCs by focusing only on the highest velocity clouds.
Furthermore, these HVCs are not entirely the compact specimens that
we believe are the most distant objects; some of the northernmost
Magellanic Stream clouds met the criteria described above, and so have
been included.  Still, many of these clouds are indeed compact HVCs,
and the combination of this list with the catalog of BB99 should
include most of the compact HVCs in the northern hemisphere.
(\citet{putman02} found 179 compact HVCs and 159 slightly more
extended clouds in the southern hemisphere in their search of the
HIPASS data, and there should be a comparable number in the northern
hemisphere.)  The positions that we searched are plotted on the sky in
Figure 2.  Finally, we point out that 12 of the BB99 HVCs and one of
the secondary clouds were also in the RB02 catalog (the others were
excluded by the velocity and declination restrictions).  In these
cases, we searched around the central positions given by each catalog,
so there was necessarily some overlap.  Several of the RB02 HVCs were
also separated from each other by less than $1\degr$.  Taking such
occurrences into account, our survey covers 264 unique HVCs and
approximately 239 degrees$^{2}$ of sky.  We also searched 54
degrees$^{2}$ in areas that do not contain an HVC, which can be used
for a comparison of results.

\subsection{Sky Survey Images}

STScI has made available digitizations of the following sky surveys
that are of use to us\footnote{Note that this description refers to
the time of the beginning of our search (late 1999); at present, the
DSS includes almost all of the POSS-II red data, and much of the
POSS-II blue data as well.}: POSS-I\footnote{The National Geographic
Society - Palomar Observatory Sky Atlas (POSS-I) was made by the
California Institute of Technology with grants from the National
Geographic Society.} (red plate only) for the northern hemisphere;
POSS-II\footnote{ The Second Palomar Observatory Sky Survey (POSS-II)
was made by the California Institute of Technology with funds from the
National Science Foundation, the National Aeronautics and Space
Administration, the National Geographic Society, the Sloan Foundation,
the Samuel Oschin Foundation, and the Eastman Kodak Corporation. The
Oschin Schmidt Telescope is operated by the California Institute of
Technology and Palomar Observatory.} (red plate only) for
approximately 2/3 of the northern sky; SERC-J\footnote{The UK Schmidt
Telescope was operated by the Royal Observatory Edinburgh, with
funding from the UK Science and Engineering Research Council (later
the UK Particle Physics and Astronomy Research Council), until 1988
June, and thereafter by the Anglo-Australian Observatory. The blue
plates of the southern Sky Atlas and its Equatorial Extension
(together known as the SERC-J), as well as the Equatorial Red (ER),
and the Second Epoch [red] Survey (SES) were all taken with the UK
Schmidt.} plates (blue) for the southern sky; Second Epoch Southern
Sky Survey (red) for a small fraction of the southern hemisphere; and
Equatorial Red plates for the region around $\delta = 0\degr$.  This
means that at any position we only have access to images of one color,
except for a small band around the equator.  Furthermore, many
positions are only covered by a single plate (especially at $\delta
\le -3\degr$) from one of these five surveys, making it difficult to
distinguish the numerous plate flaws and background variations from
real astronomical objects.

By examining the existing data on HI in known dSph galaxies, we
determined that in most cases the centroid of the HI distribution is
less than $30\arcmin$ from the optical galaxy.  Of the 10 dSphs found
by \citet{br00} to contain \hi\ gas, 7 meet this criterion, and
therefore would have been discovered in our survey if they had not
already been known.  Extending the search radius to $1\degr$ would
guarantee that we would find virtually all dSphs that are similar to
the known ones, but we judged that the $400\%$ increase in area to
search that this would require was not worth the $\sim40\%$ gain in
sensitivity.  Thus, in our survey for new galaxies, we searched a box
$1\degr$ on a side centered on each HVC in the sample.  For the BB99
compact HVCs, we used the POSS-I and SERC-J survey data, because
they had the most complete sky coverage.  After completing this, a
$30\arcmin$ wide box around each HVC position was re-examined using
POSS-II, SESSS, or Equatorial Red plates, as available.  This process
gave us greater sensitivity in the area near the center of each HVC.
By the time we reached the RB02 HVCs, most of the POSS-II data were
online, so we performed our search of these objects with only the
$1\degr$ POSS-II images.

The search was carried out by processing our downloaded images with
the algorithm described above.  We compared the processed frames with
the originals, blinking back and forth between them.  LSB objects
showed up as bright spots on the processed images.  Every position
where such a bright spot was seen and no obvious stellar or galactic
counterpart was visible on the unprocessed image was flagged.  After
this comparison, each flagged region was examined closely by eye on
the unprocessed image, using various stretches to best view the LSB
feature.  Some of these smudges were clearly related to bright
Galactic cirrus, other types of nebulae, plate flaws, ghost images, or
the wings of bright stars.  Such objects were eliminated from
consideration.  The remaining objects were compared to any other
frames of the field in question (POSS-I, or a neighboring POSS-II
plate) that were available from STScI.  Objects that were visible on a
second plate and appeared morphologically like a dwarf galaxy were
immediately labeled candidates.  Some objects that were not confirmed
on additional plates were also added to the candidate list if no other
plate was available for comparison, or if their morphology was
particularly convincing.

\section{POSS SEARCH RESULTS}

We surveyed one square degree of sky around each of 264 HVCs.  Because
some of the HVCs were close together, the total area searched was
about 239 degrees$^{2}$.  Our survey revealed 60 low-surface
brightness objects, for an overall candidate identification rate of
0.25 per degree$^{2}$ of sky searched.  Of these, 18 were towards the
59 BB99 compact HVCs (0.31 per degree$^{2}$) in our sample, and 5 were
in the direction of the 17 secondary objects (0.29 per degree$^{2}$).
The remaining 37 were found while examining the 188 HVCs from the RB02
catalog (0.20 per degree$^{2}$).  The positions of these LSB dwarf
galaxy candidates are given in Table 1.  For comparison, we also
searched (mostly by accident) 54 square degrees of sky that were not
associated with any known high-velocity gas, and found 10 additional
LSB features (0.19 per degree$^{2}$).  Without even following up on
the candidates, we can see that there is not a large excess of them in
the direction of HVCs relative to their abundance in random patches of
sky.  The somewhat higher smudge detection rates for the BB99 HVCs are
not statistically significant.

\section{OPTICAL FOLLOWUP OBSERVATIONS}
\subsection{Observing Setup}

We obtained followup images of each of the candidate dwarf galaxies
from Lick Observatory.  We observed with the 1m Anna L. Nickel
Telescope for 12 nights between 1999 July and 2000 April.  Conditions
during these nights were generally close to photometric, although
there were occasional clouds which our observations avoided.  The
seeing varied from 1\arcsec\ to 2.5\arcsec.  Our primary detector was
CCD5, a $1024 \times 1024$ thinned SITe CCD with $24 \mu$m pixels
($5\arcmin$ field of view).  However, for two nights during which this
instrument was unavailable, we used CCD2, an older $2048 \times 2048$
L30 464-5 thick phosphor coated CCD with $15 \mu$m pixels ($7\arcmin$
field of view).

We also used the 3m Shane Telescope for deeper followup observations.
We observed for 5 nights (2000 April 6--7 and 2000 September 28--30)
with the Prime Focus Camera.  This instrument utilizes a $2048 \times
2048$ SITe CCD with $24 \mu$m pixels to provide a $10\arcmin$ field of
view.  These nights also ranged from photometric to partly cloudy,
with variable seeing, but again, our observations were not
significantly affected by the clouds.

With both telescopes, we chose to employ a Spinrad R (\rs) filter for
most of our observations.  We avoided using standard R filters because
of the extremely strong Na D lines in the Lick Observatory sky, and we
felt that I-band observations would also result in the sky being too
bright.  The \rs\ filter is centered at $6850 \mbox{\AA}$, with a FWHM
of approximately $1500 \mbox{\AA}$ (somewhat redder, more symmetrical,
and wider than Cousins R), and was designed specifically to suppress
the nearby night sky emission lines.  \rs\ magnitudes tend to be
similar to Johnson R magnitudes --- within 0.1 mag for $-0.8 < V - R <
1.8$ \citep{djorgovski85}.  On the 1m telescope, we typically observed
each object for 1 hour, and on the 3m for 15 to 30 minutes.  During
the course of each night, we observed standard star fields chosen from
the \citet{landolt92} catalog to obtain photometric calibrations, as
described below.

\subsection{Data Reduction and Analysis}

Most of our data reduction was done in IDL, although we used
IRAF\footnote{IRAF is distributed by the National Optical Astronomy
Observatories, which is operated by the Association of Universities
for Research in Astronomy, Inc. (AURA) under cooperative agreement
with the National Science Foundation.} for some of the photometry.
Processing consisted of subtracting off a dark frame (for the 1m
images) or an overscan region (for the 3m images), flatfielding with a
twilight or dome flat, and flatfielding again with a sky flat if
necessary.  We removed cosmic rays with the qzap routine.  The frames
were then shifted and coadded to yield a single image for each object.
We obtained 8 to 30 standard star observations per night, which we
used to construct a photometric solution of the following form:
\begin{equation}
\rsm = m_{\mbox{\small{instr}}} + C + f \times (V - I) + g \times (a - 1),
\end{equation}
where $\rsm$ is the apparent Spinrad R magnitude,
$m_{\mbox{\small{instr}}}$ is the instrumental magnitude, $C$ is the
constant offset, $f$ is the color coefficient, $V - I$ is the known or
assumed Johnson-Cousins color of the object, $g$ is the tabulated
Lick mean extinction
coefficient\footnote{http://www.ucolick.org/~mountain/mthamilton/techdocs/info/lick-mean-extinct.html},
and $a$ is the airmass.  Equation 17 from \citet{djorgovski85} was
used to convert the Cousins R magnitudes that \citet{landolt92}
measured for the standard stars into Spinrad R magnitudes.  (The
diligent reader will note that Djorgovski's Equation 17 actually gives
the transformation between {\it Johnson} R and \rs.  This mismatch
between the Johnson and Cousins systems could lead to a systematic
error of order 0.1 mag, which is not large enough to affect our
results.  An equally important systematic effect is that the optical
path + detector combination we used is quite different from that which
Djorgovski calibrated in 1985.)

We used a local-maximum-finding algorithm to detect stars and galaxies
in our images.  Artificial star tests demonstrated that this algorithm
finds essentially all sources that are detected at the $10\sigma$
level or higher.  Aperture magnitudes were calculated for each of the
detected sources with the DAOPHOT package in IRAF.  These magnitudes
were converted to \rs\ using Equation 1, and assuming that $V - I =
1.1$, approximately correct for the most luminous red giants.  Since
the color coefficient was usually of order 0.1, this estimate can be
off by several tenths of a magnitude without having any appreciable
effect.

After locating the sources and deriving their magnitudes, we checked
the coadded images for overdensities of point sources around the
positions of the dwarf galaxy candidates.  We traced out a region of
interest around each candidate and compared the surface density of
sources inside and outside the regions over various magnitude bins.
We considered magnitude ranges starting with all the objects brighter
than the magnitude limit and fainter than a tip of the red giant
branch (TRGB) star at about 100 kpc: $\rsm = 16.5$ (since our survey
technique is not sensitive to galaxies closer than this distance).
We then decreased the bright limit by one magnitude at a time to
create successive bins (see Table 2 for an example).  We used Poisson
statistics to quantify the uncertainty in the number of stars expected
to lie inside the region based on the background surface density.
Overdensities of $3\sigma$ or higher were considered significant
enough to warrant further investigation.

\subsection{Results}

If there were a resolved dwarf galaxy in the field being examined, one
would expect to see an overdensity through all of the magnitude
ranges.  This overdensity would increase in significance as the bright
end of the magnitude bins approached the TRGB of the galaxy, reaching
a maximum near the bin where the TRGB was closest to the bright edge
of a bin.  Visually, a dwarf galaxy should also show an obvious
clustering of ``undetected'' faint stars which do not meet our
$10\sigma$ limit for photometry, since the red giant branch is more
populated at lower luminosities.  We searched each image for point
source overdensities of at least 3$\sigma$ in one magnitude bin.  Eight
of the counterparts met this criterion, with the most significant one
having a maximum overdensity of $\sim 11\sigma$.

As a comparison, we observed the dwarf spheroidals And III and And V.
And III has a total luminosity of $M_{V} = -10.2$ and a
central surface brightness of $\sbv = 24.49$ \surfb, and given its
distance of 760 kpc and the Galactic extinction ($A_{R} = 0.15$ mag),
the TRGB should be located at $R = 21.15$
\citep{caldwell92,armandroff93,sfd98}.  And V is a significantly
fainter galaxy, with $M_{V} = -9.1$ and $\sbv = 25.01$
\surfb, and is slightly farther away and more extincted (810 kpc,
$A_{R} = 0.33$ mag), so it should show a red giant branch starting at
$R = 21.48$ \citep{caldwell99,a98,sfd98}.  Our results for 3m
observations of And III (15 minutes) and And V (21 minutes) are given
in Table 2.  Based on a comparison between the stellar densities and
the background surface density, we detected about 160 stars in And III
and 80 in And V.  Both galaxies were detected at the $10.9\sigma$ level
or higher in every magnitude bin, and the highest overdensities ($\sim
15\sigma - 20\sigma$) appear in the bins that begin closest to the
TRGB magnitudes, matching our expectations.  The decreased
overdensities seen in the last bin imply that the observations were
affected to some degree by crowding or incompleteness at that
magnitude level.  We also imaged And V for 20 minutes with the 1m
telescope (this exposure was much shorter than any of our observations
of dwarf candidates) and detected the galaxy at 4.0$\sigma$ in the
faintest magnitude bin.  This significance level corresponded to 12
stars above the background density over the area of the dwarf galaxy.

Since we detect known dwarf galaxies very easily with typical 3m
observations, and weakly with much shorter than average 1m
observations, we can be confident that our observations were
sufficient to locate any new galaxies similar to the known ones.
Because the strongest candidate detections we made are still weaker
than And V --- one of the least luminous galaxies known --- we also
know immediately that if there are any dwarf galaxies associated with
HVCs, they are fainter than known dwarfs.

We also found that 15 of the dwarf galaxy candidates appeared to
contain faint nebular emission in our followup images.  We suspect
that a number of these are Galactic cirrus clouds.  The remaining 37
candidates were not detected in any way during the followup
observations, so we are forced to assume that these represent flaws on
the Palomar plates which happened to have the appearance of dwarf
galaxies.

\subsection{Comments on Individual Candidates}

{\bf HVC 104$-$70$-$312:} At the location of this candidate, we found
a $5\sigma$ density enhancement of faint sources, which contains 27
more objects than expected from the background, and peaks in
significance around magnitude 19.3.  It is higher than $4.5\sigma$ in
all but the faintest magnitude bin.  The sources appear to be
clustered around a bright ($\rsm = 18.9$) galaxy near the middle of
the region.  In our Lick image, it is obvious that most of the sources
are within 1\arcmin\ of this galaxy, rather than occupying the full
$4\arcmin \times 2\arcmin$ LSB area identified on the POSS plate.
Taking this into account, the overdensity increases in significance to
$\sim9\sigma$ (21 more objects than expected).  Of the sources
brighter than $\rsm = 20.4$ (2 magnitudes brighter than the detection
limit), 5 out of 9 are classified as galaxies on the basis of their
radial profiles (FWHM larger than the point-spread function).  For the
fainter sources, it rapidly becomes impossible to distinguish stars
from galaxies, but at least half of the 16 objects between $\rsm =
20.4$ and $\rsm = 21.4$ appear to be stars.  So there may be a group
of $15 - 20$ faint ($\rsm \ge 20.4$) stars at this position.  There
are also $\sim20$ objects within 1\arcmin\ of the central galaxy which
are visible by eye, but are fainter than our software detection limit.
Without further observations, we cannot rule out the possibility that
this is an exceedingly dim dwarf galaxy ($M_{V} \approx -7$), but we
suspect that other alternatives (e.g., galaxy cluster, random grouping
of foreground stars) are more likely.

{\bf HVC 118$-$58$-$373:} Most of the sources associated with this
small (11 objects) overdensity appear to be stars.  The significance
stays almost constant (as does the overabundance by number) down to
the $20.6-22.6$ bin, where it reaches a maximum of $3.5\sigma$.
However, even with the fairly deep limiting magnitude of this image,
there is no hint of numerous fainter stars in the region below our
formal $10\sigma$ detection limit, which should of course be present
if this were a dwarf galaxy.  With so few objects here the statistics
are obviously poor, but we argue that this candidate is probably a
chance alignment of foreground stars.

{\bf HVC 171$-$54$-$235:} The overdensity is highest ($3.6\sigma$) in
the largest magnitude range and decreases monotonically thereafter.
This behavior is not at all what is expected from a dwarf galaxy.
Again, there is no sign of ``undetected'' ($< 10\sigma$) sources.
Since the overabundant objects are bright and are not accompanied by
more numerous faint counterparts, this candidate is not a dwarf
galaxy.

{\bf HVC 237$+$50$+$078:} This candidate has an overdensity of 25
stars, which peaks at a significance level of $3.1\sigma$ around
magnitude 18.  The sources are not very centrally concentrated;
rather, many of them are in several small groups of about 6 objects.
Of the fraction (about 2/3) that are bright enough to be solidly
classified as either stars or galaxies based on their radial profiles,
over $60\%$ are galaxies.  Thus, the overdensity of {\it stars} is
small, and there is no reason to suspect that this is a dwarf galaxy.

{\bf HVC 261$+$49$+$160:} This object was the first distant galaxy
cluster that we accidentally discovered.  We noticed a highly
significant ($7.8\sigma$) clustering of ``stars'' in our Lick 3m
image.  Comparison of the SERC-J and Equatorial Red plates confirmed
that these objects had red colors.  However, the fact that the
brightest of the stars appeared to have slightly extended radial
profiles relative to the point-spread function of the Lick image
raised suspicions that they were actually marginally resolved distant
galaxies.  G. Illingworth and V. Tran acquired an image of the cluster
with the Low Resolution Imaging Spectrometer (LRIS) \citep{oke95} on
the Keck\footnote{The W.M. Keck Observatory is operated as a
scientific partnership among the California Institute of Technology,
the University of California, and the National Aeronautics and Space
Administration. The Observatory was made possible by the generous
financial support of the W.M. Keck Foundation.} I telescope which
supported this interpretation (see Figure 3), so we obtained spectra
of the brighter objects with the Hobby-Eberly Telescope\footnote{The
Hobby-Eberly Telescope is a joint project of the University of Texas
at Austin, Pennsylvania State University, Stanford University,
Ludwig-Maximillians-Universitat Munchen, and Georg-August-Universitat
Gottingen.}.  The spectra revealed that the bright central sources in
this field were compact elliptical galaxies at $z=0.35$.  Thus, we
conclude that this is a cluster of galaxies.  (We should point out
that technically this position was not listed in either of the HVC
catalogs we used; it was only examined by mistake.  However, there is
actually high-velocity \hi\ emission at this position in the LDS that
was too weak to be included in the RB02 catalog.  This cloud is also
listed in the compilation of \citet{wvw}, so it really is an HVC.)

{\bf HVC 143$+$65$+$285b:} This candidate also appeared in a 3m image
as a small clustering of slightly extended objects (23 more than would
be expected from the background surface density).  We did not obtain
spectra, but I-band imaging from the 1m telescope confirmed that many
of these objects have similar colors.  The significance of the
clustering reaches a maximum of $3.9\sigma$ and begins to fall off
around magnitude 19.  $64\%$ of the bright objects in the region are
clearly galaxies, and a 15th magnitude star is also present.  These
probably combine to explain the source of the LSB emission our
algorithm picked up on the POSS plates.  We consider this candidate to
be a likely galaxy cluster.

{\bf HVC 083$-$50$-$325:} The morphology of this candidate is
extremely suggestive of a dwarf galaxy: an elliptical region about
5\arcmin\ across with a very strong concentration of faint point
sources (Figure 4).  This was one of the two objects whose
overdensities approached those of And III and And V in significance,
peaking at $10.8\sigma$.  However, a number of the brightest objects
are clearly galaxies in our Lick \rs\ image, and a significant
fraction of those appear to be interacting.  We also obtained V and I
images of this field to aid in classifying the sources in the region
of enhanced surface density.  Taking into consideration the radial
profile, morphology, and color of each source, we classified 62\% of
the identifiable objects as galaxies.  Many of the galaxies were
concentrated in color-color space around ($V-I$, $R-I$) = (2.0, 1.1)
and also near (1.5, 0.7).  Furthermore, an Abell cluster (2545) with a
photometric redshift of 0.17 \citep{gal} is located 20\arcmin\ away
from this position, increasing the probability that other clusters are
in the vicinity.  We feel safe in concluding that this candidate is
another cluster of galaxies at moderate redshift.

{\bf HVC 107$-$30$-$421:} We found a small ($\sim 1\arcmin$),
elongated, very faint nebula north of the center of this HVC on the
Palomar plates.  A deep exposure on the 1m telescope confirmed the
reality of the source, but its extremely low surface brightness ($\sbr
\approx 25.3$ \surfb) made it difficult to classify.  A. Bunker,
S. Dawson, A. Dey, H. Spinrad, and D. Stern obtained an R-band image
for us with Keck/LRIS, which revealed more structure in the object,
but definitively did not resolve it into stars (see Figure 5).  We
remain unsure of the nature of this object, but we believe that it is
a Galactic nebula.  Because no stars were visible in it in the deep,
subarcsecond-seeing image from Keck, we are confident that it is not a
Local Group dwarf galaxy.

{\bf HVC 114$-$11$-$441:} This candidate has an overdensity of
$3.3\sigma$ (39 sources) over the magnitude range $16.5 \le \rsm \le
22.8$.  The significance decreases almost monotonically in the
subsequent bins.  The Galactic extinction in this direction is
$A_{R} = 0.65$ mag \citep{sfd98}, which makes it unlikely that
a dwarf galaxy would be detected here.  Since this object is at low
Galactic latitude, the slight clustering of stars is probably located
in the foreground.

\section{DISCUSSION}

\subsection{Expected Red Giant Populations}

We were unable to find any mention in the literature of the R-band
magnitude of the TRGB.  However, we can use theoretical studies of
stellar evolution to make a reasonable estimate of its location.
\citet[hereafter G00]{girardi00} and \citet[hereafter Y01]{yi01} both
present stellar models that predict TRGB magnitudes of $M_{R} \approx
-3.4$, varying only slightly with age and metallicity within the
ranges that interest us.  For example, in their $\mbox{[Fe/H]} = -1.7$
models, Y01 find a maximum shift of 0.086 mag in the TRGB magnitude
for stars ranging in age between 4 and 15 Gyr.  G00 give the maximum
shift as 0.020 mag for the same parameters.  Likewise, in changing
metallicities at a fixed age (10 Gyr), the TRGB magnitude varies by
only 0.144 and 0.068 mag, according to Y01 and G00, respectively (over
most of the range covered by known dwarf spheroidals, $-2.3 \le$
[Fe/H] $\le -1.3$).  Given what is known about the TRGB in V and I
\citep{lfm93}, these values seem reasonable, but we can compare them
with observations as a further test.  One of the few published
datasets involving R-band observations of a large, homogeneous
population of red giants is the study of the Fornax dSph by
\citet{stetson98}.  Using their data, and an assumed distance modulus
for Fornax of 20.70 mag \citep{beau95,saviane00}, we derive a TRGB
magnitude of $M_{R} = -3.3$.  Fornax has a high metallicity for a dSph
of [Fe/H] $= -1.0$ \citep{saviane00}, so we should not be surprised
that it has a slightly fainter tip magnitude than the models give.  We
conclude that $M_{R} = -3.4$ is the best guess for the TRGB magnitude
in a typical Local Group dwarf spheroidal.

We noted in Section 4.3 that our observations of dSph companions of
Andromeda detected $\sim 100$ stars in each galaxy.  Other recent
studies of distant Local Group dwarfs have measured similar or larger
numbers of stars within one magnitude of the TRGB: $\sim100$ such
stars in And VI \citep{armandroff99}, $\sim200$ stars in And VI and
$\sim650$ stars in And VII \citep{grebel99}, 156 stars (including some
AGB contamination) in Phoenix \citep{hsm99}, and 77 stars in Tucana
\citep{shp96}.  Since typical background surface densities in our
images are $\sim 8$ stars arcminute$^{-2}$ (varying strongly with
galactic latitude) and these densities are a factor of a few higher,
such objects are easily detected.  Their absence in our survey
indicates that HVCs do not contain dwarf galaxies with typical
parameters ($M_{V} \simless -9$, $\sbv \simless 25.0$
\surfb).

\subsection{Distance and Surface Brightness Limits}

Our followup observations of dwarf galaxy candidates reached
$10\sigma$ limiting \rs\ magnitudes between 20.0 and 23.4, with a
median value of 22.2.  We can convert these limiting magnitudes into a
minimum distance at which a dwarf galaxy would have to lie in order to
have escaped detection.  If we assume that these observations must
probe 1 magnitude below the TRGB in order to detect a dwarf galaxy,
then the limiting distance is
\begin{equation}
d_{lim} = 10^{\frac{m_{lim} + 7.4}{5}} \hspace{0.06in} \mbox{pc}.
\end{equation}
For the minimum, median, and maximum limiting magnitudes we achieved,
this distance corresponds to 302 kpc, 832 kpc, and 1445 kpc,
respectively.  These distance limits are quite conservative, because
we have insisted that stars be detected at $10\sigma$ (even though
they can be visually identified in images at significantly fainter
levels), and objects as bright as the known dSphs can easily be
located on images which reach less than a magnitude below the TRGB.
Therefore, we can state with confidence that none of the candidate
dwarf galaxies are actually dwarfs within 100 -- 300 kpc of the Milky
Way, and all but two would have to be several times farther away to
have been missed.

The depth of the POSS-II portion of our search is more difficult to
assess.  The Palomar data are not of uniform sensitivity: exposure
times vary by up to a factor of two from plate to plate, and on a
single plate there is significant vignetting within $\sim 1\degr$ of
the edges.  Furthermore, different parts of the sky are covered by
different photographic surveys.  Distant LG galaxies ($\simgtr 200$
kpc) are not resolved and appear in the Palomar data as smudges with
higher surface brightnesses than the surrounding areas.  Because all
of the distant dSphs are visible by eye on the Palomar plates, we know
that the POSS sensitivity range goes at least as faint as $\sbv =
25.0$ \surfb.  Using the stellar evolution models discussed in Section
5.1 again, we estimate that these galaxies should have colors around
$V - R = 0.5$.  Thus the R-band sensitivity of the Palomar images
(since we used the red plates wherever possible) is better than $\sbr
= 24.5$ \surfb.

A few dwarf spheroidals are known to have surface brightnesses lower
than these levels.  Some of the Milky Way companions at $d < 100$ kpc
have $25.3 \le \sbv \le 26.2$ \surfb\ (i.e. Carina, Draco, Sextans,
Ursa Minor, and Sagittarius).  Out of these, Carina and Draco are
visible by eye in POSS-II images, and the others are detectable via
star counts.  Since galaxies this close are resolved into individual
stars on POSS plates, it is not entirely clear how to consider them
with regard to a limiting surface brightness.  Nevertheless, all LG
galaxies with $\sbv < 25.5$ \surfb\ are visible, and one of the two
galaxies with $\sbv = 25.5$ \surfb\ can also be seen visually.  Thus,
the naked-eye sensitivity of the red POSS-II plates is likely to be
$\sbv = 25.5$ \surfb, or $\sbr = 25.0$ \surfb.  Our processing routine
should improve these values, perhaps by $\sim 0.5$ magnitudes.  So, we
estimate that our survey is able to find galaxies down to $\sbv =
26.0$ \surfb, or $\sbr = 25.5$ \surfb.  (It is worth noting that
foreground Galactic extinction prevents us from quite reaching
these limits; the median A$_{R}$ for HVCs in our sample
is 0.18 magnitudes \citep{sfd98}.)

\subsection{Implications}

The fact that our survey did not detect \emph{any} new dwarf galaxies
toward the $\sim 250$ HVCs that we examined rules out the hypothesis
that HVCs are the gaseous components of normal, but faint, Local Group
dwarf galaxies.  If these HVCs do contain stars, they must have
central surface brightnesses $\sbv \simgtr 26$ \surfb.  They are also
likely to have absolute magnitudes $M_{V} \simgtr -9$,
although a very extended stellar counterpart with an extremely low
surface brightness could yield a higher total luminosity while still
escaping detection.  Even though it is conceivable that a few objects
could have been missed because of our survey design, on the whole it
seems clear that HVCs are starless systems.

Known LG dwarfs do have surface brightnesses as faint as $\sbv =
26.2$ \surfb, but only a handful of the very extended MW satellites
are in this range (see Figure 6). Surveys of nearby groups and
clusters confirm that dwarf spheroidals with $\sbv \simgtr 26$
\surfb\ either do not exist or cannot be detected with current
techniques
\citep{caldwell87,impey88,bothun89,bremnes98,caldwell98,jerjen00,kea00,kea01}.
For the distant LG dSphs (Andromeda companions and isolated galaxies),
the lowest surface brightness is $\sbv = 25.05$ \surfb\ for
Cetus and Tucana.  All of these distant galaxies are visible in the
POSS-II data, and are easily detected both by our POSS processing
algorithm and our followup imaging campaign.  Thus, we are confident
that any objects similar to known dwarf spheroidals would have been
discovered in our survey if they were present.  Lower surface
brightness stellar counterparts could be present, but the fact that no
such systems (with or without \hi\ components) are known argues
against this possibility.  We believe that the most likely explanation
for our findings is that HVCs simply do not contain stars.

This result does not lead to the anticipated outcome of our survey: a
means of discriminating between HVC models.  In the Galactic Fountain and
Tidal Debris models, stars would not be expected to form in HVCs, in
agreement with our finding.  However, although the Local Group model
certainly allows for stars in HVCs, and we would argue suggests that
stellar counterparts are likely, it does not require them.  So while a
positive result in our search would have strongly supported the Local
Group model, the inverse is not necessarily the case.  As mentioned
earlier, \citet{bullock00} and others have proposed ideas explaining
how small dark matter halos scattered throughout the Local Group could
contain some neutral gas without having formed stars.  We believe that
the absence of stars may be an important clue to the nature of HVCs,
but by itself, it does not allow us to solve the puzzle.

\subsection{Compatibility With Previous Work}

There have been no other comprehensive and quantitative searches for
evidence of stars in HVCs.  \citet{ic97} examined IRAS data towards
the large HVC complexes in an effort to locate any star formation that
might be occurring, but only came up with one possible young star in
positional coincidence with high-velocity \hi.  BB99 searched DSS
images in the direction of each of their compact HVCs, but found no
clear optical counterparts.  And Simon et al. (in preparation)
used infrared and millimeter-wave observations to search for a
stellar counterpart to Complex H, with null results. Thus, the result
of this work is entirely compatible with the data existing in the
literature.

\section{CONCLUSIONS}

We have surveyed one degree$^{2}$ of sky around each of 264
high-velocity clouds in search of new Local Group dwarf galaxies.  We
processed digital POSS-I and POSS-II images with an algorithm to
enhance low-surface brightness features.  We then examined the images
and found 60 faint smudges that we classified as possible LG dwarfs.
Using the 1m and 3m telescopes at Lick Observatory, we imaged each of
these candidates to a typical limiting stellar magnitude of $\rsm =
22.2$.  Examination of the data revealed several $\ge 3 \sigma$
density enhancements of faint sources in the areas selected from the
POSS plates, but none of these appear to be LG dwarf galaxies.
Typical faint LG dwarf spheroidal galaxies would have been detected at
the $\simgtr 10 \sigma$ level, with $\sim 100$ stars brighter than the
detection limit.  Therefore, we conclude that there are no
undiscovered normal dwarf galaxies within a 30\arcmin\ radius of any
of these HVCs, provided that the HVCs are located at least 100 kpc
from the Milky Way; dwarf galaxies at a smaller distance might be too
diffuse to detect in this manner.

There were both observational and theoretical grounds for suspecting
that compact HVCs might harbor faint LSB dwarf galaxies.  It is
well-known that the highly successful Cold Dark Matter theory
predicts large numbers of small dark matter halos, which have not yet
been detected observationally.  \citet{klypin99} and others have
pointed out that HVCs are numerous enough (and massive enough, in the
Local Group picture) to comprise the set of missing halos.  From the
observational side, \citet{br00} noted that there are similarities
between the \hi\ components of LG dwarfs and the properties of HVCs,
if they are located at typical distances from the Milky Way of $\sim
700$ kpc.  Furthermore, blind \hi\ surveys of nearby groups of
galaxies often find that the \hi\ clouds they detect are associated
with LSB dwarfs.

The implications of our observation that LG \hi\ clouds lack such
stellar counterparts are unclear.  While this finding does not rule
out the Local Group hypothesis for the spatial distribution of the
HVCs, it also does not provide any supporting evidence.  We are still
unable to constrain HVC distances, but we have placed
significant limits on their stellar content.  Regardless of the
location of the HVCs, as long as they are at least 100 kpc away they
either lack stars entirely, or they have lower surface brightnesses
and luminosities than other systems in the Local Group.

\acknowledgments{We acknowledge the support of grant AST-9981308.  We
thank the referee for his or her comments on the paper.  In addition,
we would like to thank the Lick Observatory staff, particularly Elinor
Gates and Tony Misch, for their instruction on the Lick telescopes.
JDS also acknowledges the following people for their company and
assistance while observing: Alison Coil, Steve Dawson, Wendy Ellis,
Doug Finkbeiner, Eric Gawiser, Tim Robishaw, and Tony Wong.  We are
also grateful to several people for being kind enough to donate some
of their own observing time to this project, specifically Andy Bunker,
Steve Dawson, Arjun Dey, Elinor Gates, Garth Illingworth, Hy Spinrad,
Dan Stern, and Vy Tran.  Frank Bash allowed us to use some of his
discretionary time on the HET to take spectra of one of our
candidates.  Doug Finkbeiner and Tim Robishaw were the sources of many
helpful ideas about IDL programming.  This research has made use of
the SIMBAD database, operated at CDS, Strasbourg, France.  This
research has also made use of the NASA/IPAC Extragalactic Database
(NED) which is operated by the Jet Propulsion Laboratory, California
Institute of Technology, under contract with the National Aeronautics
and Space Administration, and NASA's Astrophysics Data System
Bibliographic Services.}

\begin{figure}
\plotone{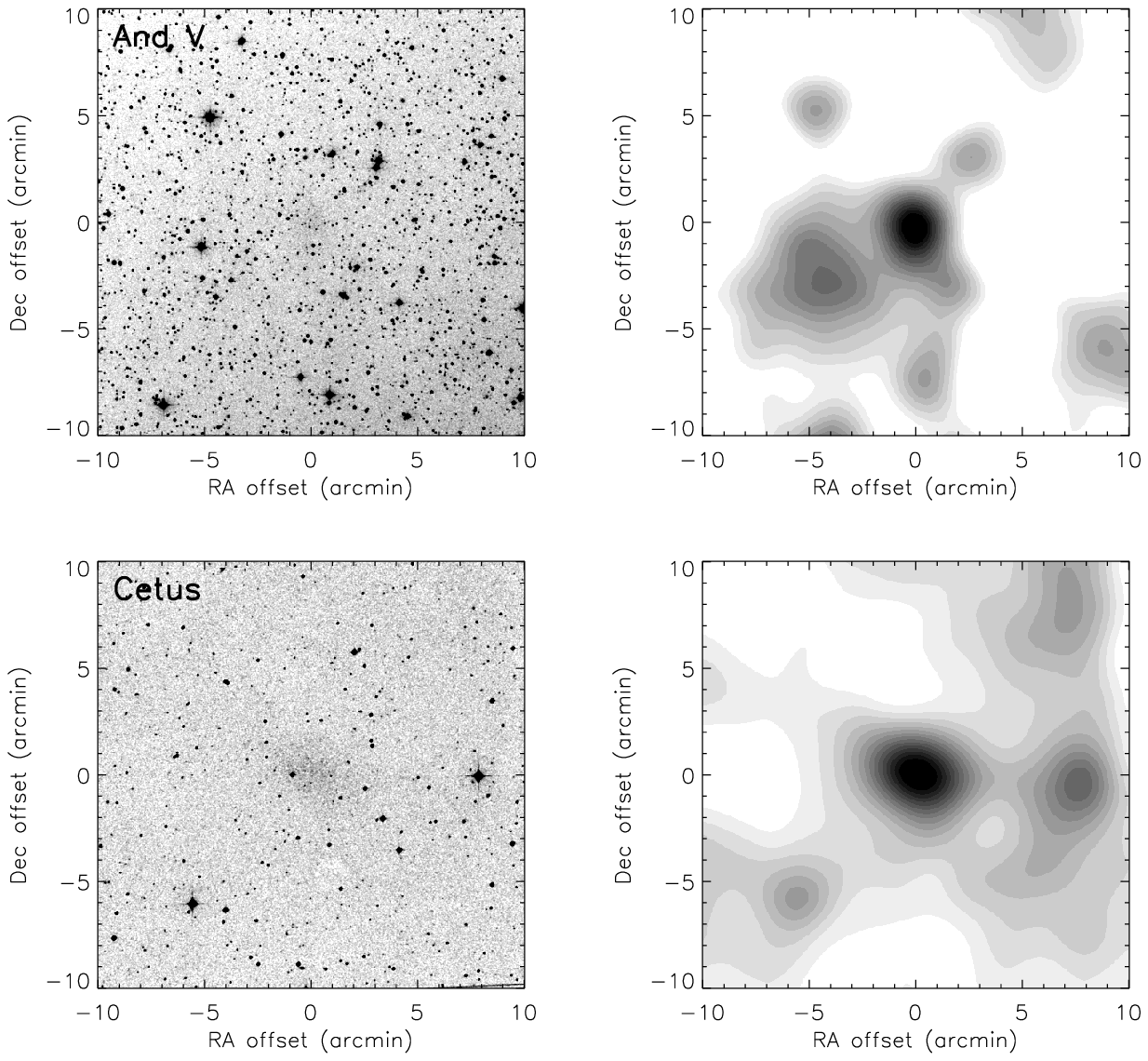}
\caption{Demonstration of our POSS processing algorithm.  On the left
are POSS-II red images of two of the lowest surface brightness dwarf
spheroidals in the Local Group: And V (top) and Cetus (bottom).  The
galaxies are just visible as faint smudges at the center of each
frame.  On the right are the same fields after applying our algorithm.
In each case, the dSph has become the brightest object in the image.}
\end{figure}

\begin{figure}
\plotone{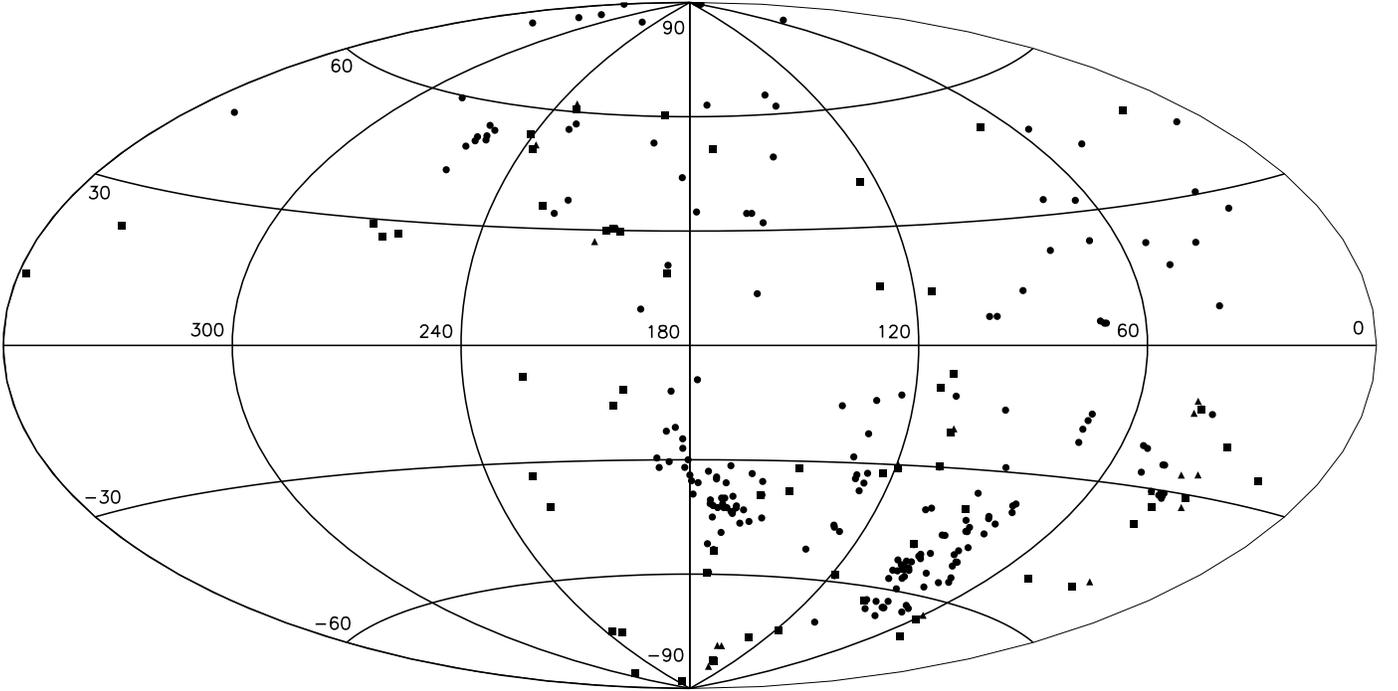}
\caption{Distribution on the sky of the HVCs we searched.  The squares
represent HVCs from the BB99 catalog, the circles are HVCs from the
catalog of Robishaw \& Blitz, and the triangles are secondary clouds
selected from BB99.  The concentration of HVCs along $\ell = 90\degr$
between $\mbox{b} = -35\degr$ and $\mbox{b} = -65\degr$ shows that
both catalogs contain significant numbers of Magellanic Stream
clouds.}
\end{figure}

\begin{figure}
\plotone{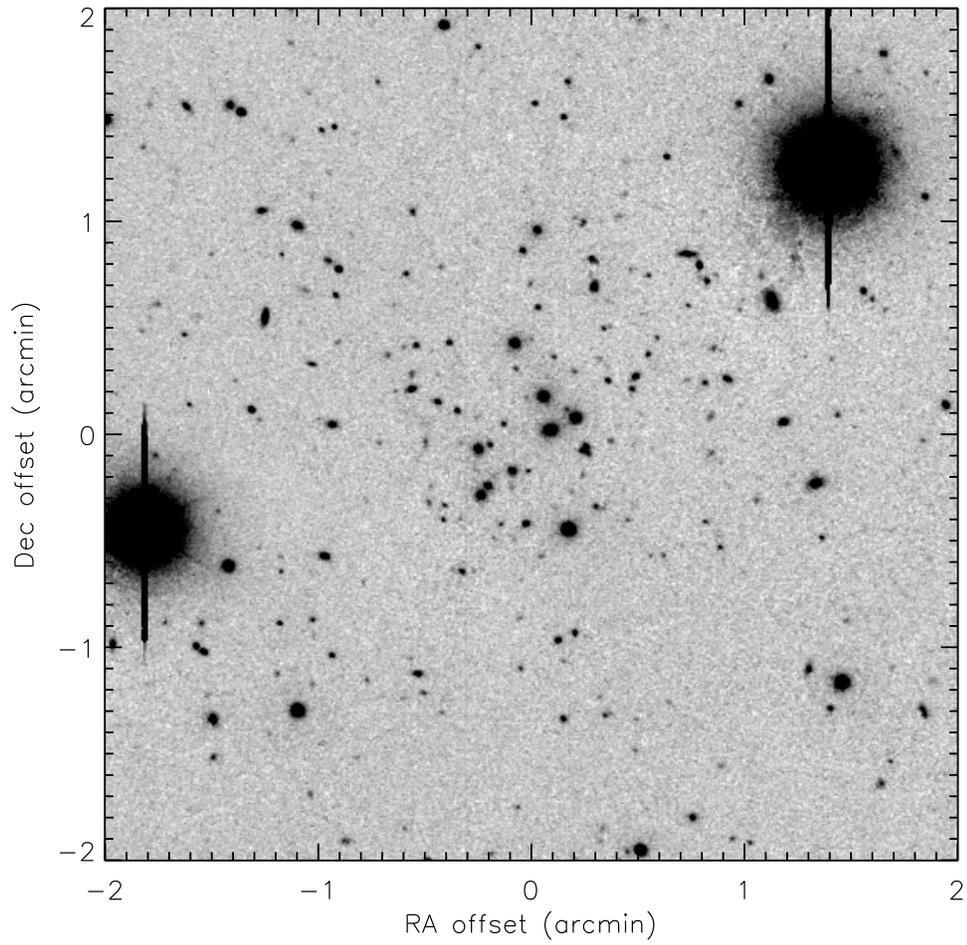}
\caption{10-minute Keck/LRIS I-band image of the dwarf galaxy candidate
261$+$49$+$160, acquired for us by G. Illingworth and V. Tran.  The
concentration of objects in the center of the frame is a
spectroscopically confirmed galaxy cluster at $z=0.35$.}
\end{figure}

\begin{figure}
\plotone{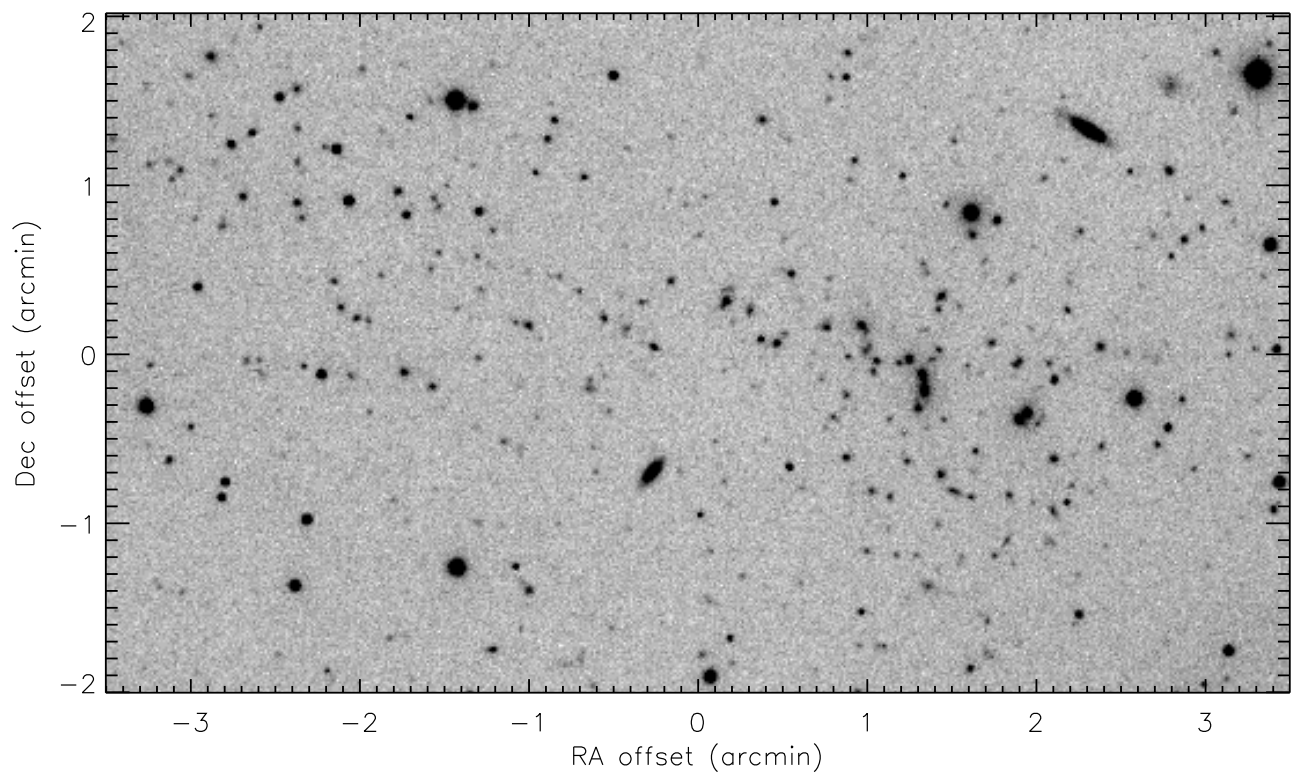}
\caption{15-minute \rs\ image from the Lick 3m telescope of the
candidate 083$-$50$-$325.  The main concentration of objects which is
described in the text includes several bright, interacting galaxies,
and is centered near (RA off, Dec off) = (1.3\arcmin, $-$0.2\arcmin).
The overdensity continues less strongly to the northeast, all the way
to the upper left corner of the image.}
\end{figure}

\begin{figure}
\plotone{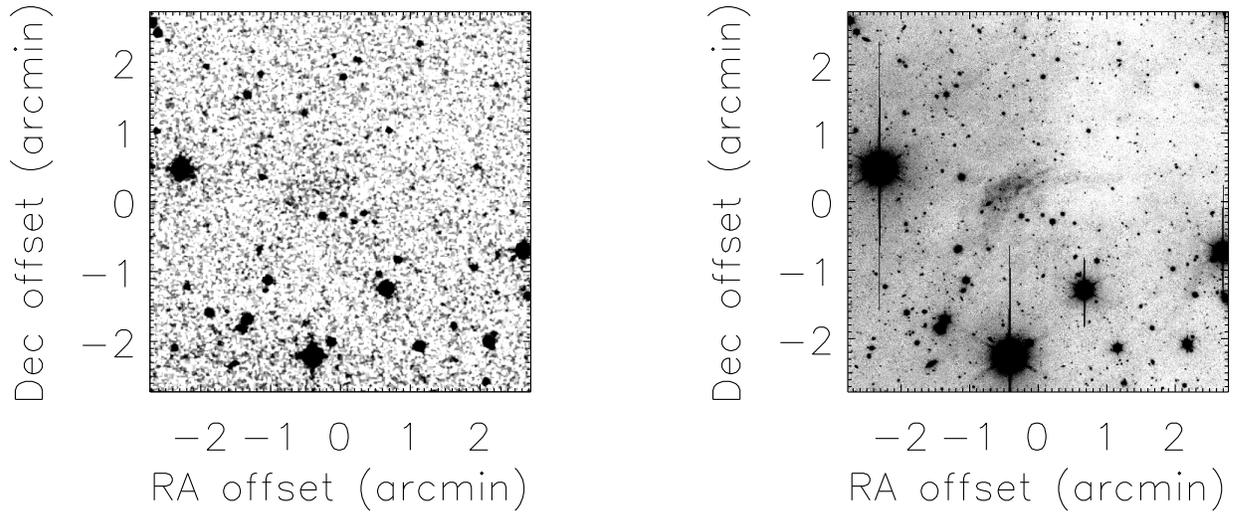}
\caption{Dwarf galaxy candidate near HVC 107$-$30$-$421.  On the left
is a POSS-II red image showing an LSB smudge just to the left of the
field center.  The R-band Keck/LRIS image of the same field, obtained
by A. Bunker and collaborators, is on the right.  Note that even
objects with surface brightnesses as low as $\sbr = 25.3$ \surfb\ are
visible by eye (with difficulty) on raw POSS-II images.}
\end{figure}

\begin{figure}
\plotone{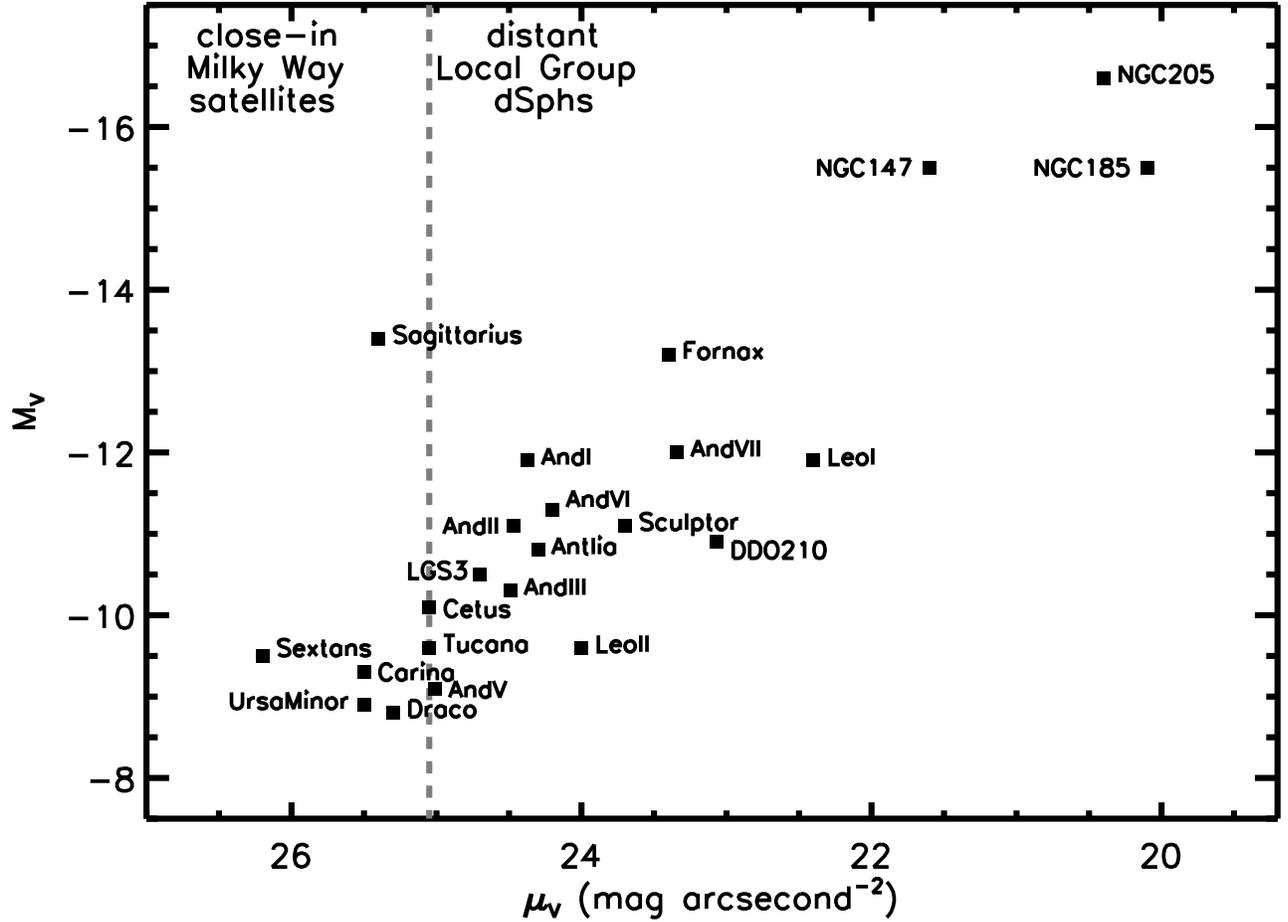}
\caption{Relationship between luminosity and surface brightness for
spheroidal galaxies, dwarf spheroidals, and the so-called transitional
(dSph/dIrr) galaxies in the Local Group (Pegasus and Phoenix are not
included in this figure because they lack published surface photometry
data).  The only known dwarfs with $\sbv > 25.05$ \surfb\ (to the left
of the dashed line) are the Milky Way satellites within 100 kpc.  All
of the M31 companions and isolated dwarfs in the LG have higher
surface brightnesses.  The data used to make this plot were taken from
tables compiled by \citet{mateo98} and \citet{vdb99,vdb00} (see
references therein for the original sources), except for the surface
photometry for DDO 210, which was adapted from \citet{lee99}.}
\end{figure}

\begin{deluxetable}{l c c c c c c }
\tabletypesize{\scriptsize} 
\tablenum{1} 
\tablewidth{0pt}
\tablecolumns{7} 
\tablecaption{The Candidate Dwarf Galaxies Identified In Our Survey} 
\tablehead{ \colhead{HVC name\tablenotemark{a}} &
\colhead{$\alpha_{\mbox{{\tiny J2000}}}$\tablenotemark{b}} &
\colhead{$\delta_{\mbox{{\tiny J2000}}}$} & \colhead{point source} &
\colhead{nebular} & \colhead{limiting \rs\ magnitude} & \colhead{HVC
catalog\tablenotemark{c}} \\ 
\colhead{} & \colhead{} & \colhead{} & \colhead{overdensity} & 
\colhead{counterpart} & \colhead{of followup image} & \colhead{}}

\startdata 
104$-$70$-$312 & 00 24 36  & $-$08 09 54   & yes        & no\phm{s}  & 22.3 & RB\phm{S} \\
120$-$30$-$289\tablenotemark{d} & 00 38 34  & \phs32 43 47  & no\phm{s}  & no\phm{s}  & 21.4 & BBS \\*
               & 00 38 40  & \phs32 48 22  & no\phm{s}  & yes        & 22.7 & BBS \\
118$-$58$-$373 & 00 42 13  & \phs04 32 07  & yes        & no\phm{s}  & 22.6 & BB\phm{S} \\
124$-$13$-$214 & 00 55 44  & \phs50 23 51  & no\phm{s}  & no\phm{s}  & 22.6 & RB\phm{S} \\
156$-$45$-$303 & 02 24 40  & \phs11 46 50  & no\phm{s}  & yes        & 21.3 & RB\phm{S} \\*
               & 02 24 59  & \phs11 21 47  & no\phm{s}  & no\phm{s}  & 21.4 & RB\phm{S} \\
171$-$54$-$235\tablenotemark{e} & 02 36 08  & $-$00 48 08 & yes & no\phm{s}  & 22.2 & BB\phm{S} \\
158$-$39$-$280 & 02 40 43  & \phs16 31 28  & no\phm{s}  & no\phm{s}  & 22.1 & RB\phm{S} \\
158$-$39$-$285 & 02 41 41  & \phs16 17 15  & no\phm{s}  & no\phm{s}  & 22.4 & BB\phm{S} \\
169$-$43$-$259 & 03 01 38  & \phs08 14 55  & no\phm{s}  & no\phm{s}  & 21.9 & RB\phm{S} \\
169$-$40$-$235 & 03 07 46  & \phs10 25 00  & no\phm{s}  & no\phm{s}  & 22.1 & RB\phm{S} \\
189$-$32$+$202\tablenotemark{f} & 04 16 23  & \phs03 47 29 & no\phm{s}  & yes & 21.5 & RB\phm{S} \\*
189$-$32$+$248 &           &               &            &            &      &    \\
182$-$25$-$203 & 04 25 34  & \phs13 26 43  & no\phm{s}  & yes        & 22.5 & RB\phm{S} \\
162$+$14$-$382 & 06 02 42  & \phs51 25 53  & no\phm{s}  & no\phm{s}  & 22.9 & RB\phm{S} \\
159$+$32$+$268 & 08 03 13  & \phs58 25 30  & no\phm{s}  & no\phm{s}  & 23.3 & RB\phm{S} \\
200$+$30$+$075 & 08 24 04  & \phs23 38 59  & no\phm{s}  & yes        & 23.0 & BB\phm{S} \\
204$+$30$+$061 & 08 26 44  & \phs20 15 15  & no\phm{s}  & no\phm{s}  & 22.3 & BB\phm{S} \\
237$+$50$+$078 & 10 24 49  & \phs06 36 41  & yes        & no\phm{s}  & 22.7 & BB\phm{S} \\
237$+$51$+$120 & 10 30 23  & \phs07 47 05  & no\phm{s}  & no\phm{s}  & 22.6 & BBS \\
261$+$49$+$160 & 11 07 16  & $-$04 54 41   & yes        & no\phm{s}  & 22.6 & RB\phm{S} \\
143$+$65$+$285 & 12 00 44  & \phs50 06 15  & no\phm{s}  & no\phm{s}  & 22.8 & RB\phm{S} \\*
               & 12 02 51  & \phs50 22 58  & yes        & no\phm{s}  & 23.2 & RB\phm{S} \\
050$+$81$-$442 & 13 29 31  & \phs29 13 00  & no\phm{s}  & no\phm{s}  & 23.4 & RB\phm{S} \\
347$+$46$-$215 & 14 49 56  & $-$06 42 22   & no\phm{s}  & no\phm{s}  & 21.9 & RB\phm{S} \\*
347$+$46$+$259 &           &               &            &            &      &    \\
070$+$51$-$146 & 15 48 55  & \phs43 49 50  & no\phm{s}  & no\phm{s}  & 21.9 & BB\phm{S} \\
046$+$44$+$201\tablenotemark{g} & 16 20 30  & \phs26 54 03  & no\phm{s}  & no\phm{s}  & 22.3 & RB\phm{S} \\
068$+$24$-$269 & 18 14 17  & \phs40 53 18  & no\phm{s}  & yes        & 23.2 & RB\phm{S} \\
080$+$22$-$205 & 18 44 47  & \phs50 31 58  & no\phm{s}  & yes        & 23.2 & RB\phm{S} \\
043$-$13$-$314 & 19 55 56  & \phs02 43 48  & no\phm{s}  & yes        & 22.1 & BB\phm{S} \\*
043$-$13$-$267 &           &               &            &            &      &    \\
039$-$27$-$310 & 20 36 19  & $-$07 11 15   & no\phm{s}  & no\phm{s}  & 21.9 & BBS \\
046$-$25$-$235 & 20 41 02  & $-$01 16 49   & no\phm{s}  & no\phm{s}  & 21.4 & RB\phm{S} \\
031$-$33$-$330 & 20 47 33  & $-$16 20 36   & no\phm{s}  & no\phm{s}  & 21.6 & BBS \\
039$-$31$-$278 & 20 51 55  & $-$08 26 19   & no\phm{s}  & no\phm{s}  & 21.9 & BB\phm{S} \\*
               & 20 52 21  & $-$08 17 35   & no\phm{s}  & yes        & 21.8 & BB\phm{S} \\
040$-$31$-$272 & 20 53 22  & $-$08 25 30   & no\phm{s}  & no\phm{s}  & 21.6 & RB\phm{S} \\
043$-$31$-$236 & 20 58 39  & $-$05 41 00   & no\phm{s}  & no\phm{s}  & 21.8 & RB\phm{S} \\
072$-$16$-$395 & 21 10 03  & \phs24 01 16  & no\phm{s}  & no\phm{s}  & 20.2 & RB\phm{S} \\
039$-$37$-$238 & 21 15 21  & $-$11 40 21   & no\phm{s}  & no\phm{s}  & 21.8 & BB\phm{S} \\
079$-$37$+$213 & 22 32 37  & \phs13 10 41  & no\phm{s}  & no\phm{s}  & 22.4 & RB\phm{S} \\*
079$-$37$+$235 &           &               &            &            &      &    \\*
079$-$37$+$252 &           &               &            &            &      &    \\
080$-$42$-$329 & 22 46 29  & \phs10 09 49  & no\phm{s}  & yes        & 22.2 & RB\phm{S} \\
083$-$49$-$307 & 23 09 23  & \phs05 27 13  & no\phm{s}  & no\phm{s}  & 21.1 & RB\phm{S} \\
083$-$50$-$325 & 23 11 24  & \phs05 17 13  & yes        & no\phm{s}  & 22.2 & RB\phm{S} \\
050$-$68$-$201 & 23 23 12  & $-$19 09 33   & no\phm{s}  & no\phm{s}  & 21.1 & BB\phm{S} \\
111$-$07$-$466 & 23 24 53  & \phs53 58 00  & no\phm{s}  & no\phm{s}  & 22.6 & BB\phm{S} \\
               & 23 26 30  & \phs53 49 12  & no\phm{s}  & yes        & 22.8 & BB\phm{S} \\
093$-$52$-$312 & 23 38 57  & \phs06 09 19  & no\phm{s}  & no\phm{s}  & 21.1 & RB\phm{S} \\*
093$-$52$-$266 &           &               &            &            &      &    \\
108$-$21$-$395 & 23 39 46  & \phs39 25 39  & no\phm{s}  & no\phm{s}  & 20.0 & BB\phm{S} \\*
               & 23 40 37  & \phs39 23 25  & no\phm{s}  & no\phm{s}  & 21.3 & BB\phm{S} \\
093$-$55$-$276 & 23 43 02  & \phs04 28 13  & no\phm{s}  & no\phm{s}  & 21.5 & RB\phm{S} \\
080$-$66$-$226 & 23 44 55  & \phs09 15 57  & no\phm{s}  & no\phm{s}  & 22.1 & RB\phm{S} \\
097$-$53$-$384 & 23 48 35  & \phs07 07 42  & no\phm{s}  & yes        & 21.5 & RB\phm{S} \\
107$-$30$-$421 & 23 48 55  & \phs31 28 44  & no\phm{s}  & yes        & 22.6 & BB\phm{S} \\
114$-$11$-$441 & 23 49 28  & \phs50 57 13  & yes        & no\phm{s}  & 22.8 & RB\phm{S} \\
100$-$49$-$395 & 23 50 19  & \phs10 48 33  & no\phm{s}  & no\phm{s}  & 22.9 & RB\phm{S} \\*
               & 23 50 51  & \phs11 27 13  & no\phm{s}  & no\phm{s}  & 23.0 & RB\phm{S} \\
097$-$54$-$363 & 23 51 04  & \phs05 54 15  & no\phm{s}  & no\phm{s}  & 22.4 & RB\phm{S} \\
096$-$58$-$273 & 23 54 18  & \phs02 36 42  & no\phm{s}  & yes        & 21.5 & RB\phm{S} \\
114$-$10$-$440 & 23 55 04  & \phs51 26 45  & no\phm{s}  & yes        & 21.1 & BB\phm{S} \\
089$-$65$-$312 & 23 55 35  & $-$06 04 29   & no\phm{s}  & no\phm{s}  & 21.5 & RB\phm{S} \\

\enddata 

\tablenotetext{a}{HVC names are given using the convention of
\citet{bb99}: 3-digit Galactic longitude, appended with 2-digit
Galactic latitude and 3-digit LSR velocity.}

\tablenotetext{b}{The candidate positions should be accurate to 
$\approx30\arcsec$.}

\tablenotetext{c}{BB refers to HVCs from the BB99 catalog, BBS to
secondary clouds selected from BB99's moment maps, and RB to HVCs from
the RB02 catalog.}

\tablenotetext{d}{In several cases, we found two LSB objects near one
HVC; for these HVCs, the LSB counterparts are listed in order of
increasing RA.  If we have to refer to them by name, they are called
$\ell\ell\ell-bb-vvv$a and $\ell\ell\ell-bb-vvv$b.}

\tablenotetext{e}{Note that BB99 list the wrong declination for this
HVC in their Table 1.  The actual declination for the given galactic
coordinates is $-$00 55, not +00 55.}

\tablenotetext{f}{There are multiple HVCs at this position, so the
individual components are listed on separate lines.  The same is true
for five other HVCs in the table.}

\tablenotetext{g}{We observed this HVC in \hi\ with the 305m Arecibo
radio telescope for a separate project and did not detect it, despite
the clear signal present at this position in the LDS.  We speculate
that the LDS ``detection'' might be interference and that there may
not actually be an HVC here.}

\end{deluxetable}

\begin{deluxetable}{l c c l c c}
\tablenum{2}
\tablewidth{0pt}
\tablecolumns{6}
\tablecaption{Detections of Stars in Known Local Group Dwarf Spheroidals}
\tablehead{
\colhead{Galaxy} & \colhead{Magnitude Range} & \colhead{Stellar Overdensity} & 
\colhead{Galaxy} & \colhead{Magnitude Range} & \colhead{Stellar Overdensity}}

\startdata 
And III & $16.5 - 23.0$ & $18.8\sigma$ & And V & $16.5 - 22.9$ & $10.9\sigma$ \\
        & $17.0 - 23.0$ & $18.8\sigma$ &  & $16.9 - 22.9$ & $10.9\sigma$ \\
        & $18.0 - 23.0$ & $19.3\sigma$ &  & $17.9 - 22.9$ & $11.3\sigma$ \\
        & $19.0 - 23.0$ & $19.9\sigma$ &  & $18.9 - 22.9$ & $12.0\sigma$ \\
        & $20.0 - 23.0$ & $20.7\sigma$ &  & $19.9 - 22.9$ & $12.8\sigma$ \\
        & $21.0 - 23.0$ & $21.8\sigma$ &  & $20.9 - 22.9$ & $14.6\sigma$ \\
        & $22.0 - 23.0$ & $11.5\sigma$ &  & $21.9 - 22.9$ & $13.2\sigma$ \\

\enddata
\end{deluxetable}

\end{document}